\documentclass[review]{elsarticle}

\usepackage{lineno,hyperref}
\hypersetup{colorlinks = true,linkcolor = blue,anchorcolor =red,citecolor = blue,filecolor = red,urlcolor = red}
\usepackage{gensymb}
\usepackage{upgreek}
\usepackage{url}
\usepackage{hyperref}
\usepackage{bm}
\usepackage{nicematrix}
\journal{Journal of \LaTeX\ Templates}

\newcommand{\RNum}[1]{\uppercase\expandafter{\romannumeral #1\relax}}
\begin{document}

\begin{frontmatter}

\title{Performance of a focal plane detector for soft X-ray imaging spectroscopy based on back-illuminated sCMOS}

% \address{Radarweg 29, Amsterdam}
% \fntext[myfootnote]{Since 1880.}

\author[a,b]{Can Chen}
\author[a]{Yusa Wang \corref{mycorrespondingauthor}}\cortext[mycorrespondingauthor]{Corresponding author. \\ \indent ~~Email address: \href{wangyusa@ihep.ac.cn}{wangyusa@ihep.ac.cn}}
\author[a,b]{Yupeng Xu}
\author[a]{Zijian Zhao}
\author[c]{Hongyun Qiu}
\author[a]{Dongjie Hou}
\author[a]{Xiongtao Yang}
\author[a]{Jia Ma}
\author[a]{Yong Chen}
\author[c]{Yang Zhao}
\author[c]{Hua Liu}
\author[a]{Xiaofan Zhao}
\author[a,d]{Yuxuan Zhu}

\address[a]{Key Laboratory of Particle Astrophysics, Institute of High Energy Physics, Chinese Academy of Sciences, Beijing 100049, China.}
\address[b]{University of Chinese Academy of Sciences, Chinese Academy of Sciences, Beijing 100049, China.}
\address[c]{QHYCCD Ltd, Beijing 100084, China.}
\address[d]{College of Physics, Jilin University, Changchun 130012, China}

% e-mail addresses: only for the corresponding author
% \emailAdd{wangyusa@ihep.ac.cn}
% \emailAdd{xuyp@ihep.ac.cn}

\begin{abstract}
Spectroscopy focusing array (SFA) and Polarimetry focusing array (PFA) are the two major payloads of enhanced X-ray Timing and Polarimetry mission (eXTP). Nested Wolter-\RNum{1} X-ray mirror module is implemented in SFA and PFA to achive high effective area. 
When evaluating the properties of the mirror module, the alignment of the optical axis of the X-ray mirror module and a quasi-parallel X-ray beam is a prerequisite to ensure the accuracy of the results. 
Hence, to assist the alignment of the X-ray mirror module, a X-ray focal plane detector is designed based on the back-illuminated scientific Complementary Metal-Oxide-Semiconductor Transistor (sCMOS) sensor GSENSE6060BSI, one of the largest detection areas, is produced by \textit{Gpixel Inc}.  
Then the characteristics of readout noise, dark current, and split-pixel event properties of the detector are studied with the self-developed multi-target fluorescence X-ray source in a 100 m long X-ray test facility. 
The energy calibration is carried out with the single-pixel event and the energy non-linearity of the detector is also obtained. 
% Additionally, X-ray imaging experiment with patterned mask and simulation based on optical model are conducted to confirm the imaging capability of the detector.
Eventually, the simulation of the eXTP mirror module based on the optical model is conducted and the alignment test of the Wolter-\RNum{1} X-ray mirror module designed for \textit{EP/FXT} (Einstein Probe/Follow-up X-ray Telescope) with ``Burkert test'' method is shown.
\end{abstract}

\begin{keyword}
Back-illuminated CMOS sensor; X-ray detector; X-ray optics; Alignment; Imaging spectroscopy; eXTP
% X-ray detectors; Solid state detectors; Vacuum-based detectors; Imaging spectroscopy
\end{keyword}

\end{frontmatter}

%\linenumbers

\section{Introduction}
Enhanced X-ray Timing and Polarimetry mission (eXTP), expected to be launched in 2027, is an international cooperation project led by the Institute of High Energy Physics (IHEP), Chinese Academy of Science (CAS), whose scientific goal is to study the state of matter under extreme conditions \cite{zhang2019enhanced}. 
Spectroscopy focusing array (SFA), Polarimetry focusing array (PFA), Large area detector (LAD), and Wide field monitor (WFM) are the four main payloads of eXTP. 
Wolter-\RNum{1} type X-ray mirror module is adopted in SFA and PFA to achive high effective area. 

The Wolter-\RNum{1} telescope is composed of a coaxial confocal internal reflecting paraboloid and an internal reflecting hyperboloid. The focal point of the paraboloid coincides with the focal point of the hyperboloid. 
The light incident on the internal reflecting surface of the parabola and should be focused after reflection to its focal point, but after being reflected twice by the internal reflecting surface of the hyperboloid, it converges to another focal point of the hyperboloid \cite{wolter1975mirror}. 
The shape of the single reflected light from the paraboloid or hyperboloid surface on the focal plane is sensitive to the pitch and yaw angles of the Wolter-\RNum{1} telescope to the incident optical axis.
Therefore, the reflection characteristics of the mirror are used to find a rough range for further precise focus position.
This fast alignment method that provides good pitch and yaw angles scanning range is called the ``Burkert test'' \cite{BenediktMenz2013}\cite{bradshaw2019developments} by the PANTER, an X-ray test facility located in Munich, Germany.
% The initial estimation of the angle has the advantage of dramatically reducing the time required for the pitch and yaw angles scanning during fine scanning. 

The focal length of the mirror module adopted in eXTP  is 5.25 m, and the observation X-ray energy range is from 0.5 to 10 keV \cite{S.Bassoextp2019}. 
According to the result of the optical simulation, an imaging detector with an area larger than 60 mm $\times$ 60 mm is able to meet the requirements of the ``Burkert test'' method, seen in section \ref{X-ray imagination}.

Charge couple device (CCD) and CMOS (Complementary Metal-Oxide-Semiconductor Transistor) are commonly used as focal plane detector in soft X-ray imaging for their superior energy and spatial resolution. 
However, the tailing effect in CCD is obvious due to the charge transfer process when there is no shutter in front of the sensor \cite{898423}.
Besides, the advantages of relatively simple peripheral circuit design have made CMOS gradually applied to many fields \cite{bigas2006review}\cite{ROGALSKI2012342}, including X-ray imaging polarimetry \cite{asakura2019x} or even as a focal plane detector for some micro or small X/$\upgamma$-ray scientific mission \cite{nakajima2020developing}\cite{SteveTammes2020}. Therefore, the CMOS sensor is selected as the X-ray focal plane detector to test the X-ray mirror module. 

Back-illuminated sCMOS sensor can achieve high quantum efficiency and low noise in soft X-ray imaging spectroscopy.
GSENSE400BSI and GSENSE6060BSI are two ``GSENSE'' series sCMOS sensors produced by \textit{Gpixel Inc.} 
The former has a wide range of applications \cite{desjardins2019characterization}\cite{Wang_2019}\cite{Ling_2021} due to its high cost performance and mature technology.
The latter is one of the largest detection areas sCMOS sensor produced by \textit{Gpixel Inc.}, which meets the demand of the ``Burkert test'' method in the detection area. However, there is no relevant research on the GSENSE6060BSI, especially under high vacuum (about $10^{-5}$ Pa) and low temperature (below -30 $^{\circ}$C) condition. 
The specification of GSENSE6060BSI is shown in Table~\ref{specification of GSENSE6060BSI}.

\begin{table}\label{tab_specification_6060}
    \centering
    \caption{Specification of GSENSE6060BSI \cite{GSENSE6060BSI_datasheet}}
    \label{specification of GSENSE6060BSI}
        \begin{tabular}{cc}
        \hline
        Items                & Description                    \\ \hline
        Resolution           & 6144~$\times$~6144               \\ \hline
        Pixel size           & $10~{ \upmu}\rm{m}\times{10~\upmu \rm{m}}$   \\ \hline
        Photonsensitive area & 61.44 mm~$\times$~61.44 mm       \\ \hline
        ADC                  & 14 bit                         \\ \hline
        Shutter type         & Rolling shutter (global reset) \\ \hline
        Operation Mode & \begin{tabular}[c]{@{}c@{}}14 bit Standard\\ 12 bit Standard\\ 12 bit HDR\end{tabular}  \\ \hline
        Output interface & \begin{tabular}[c]{@{}c@{}}50 $\times$ LVDS @ 420 Mbps (12 bit)\\ 14 $\times$ LVDS @ 420 Mbps (14 bit)\end{tabular} \\ \hline
        Dark current         & \begin{tabular}[c]{@{}c@{}}20 $\rm e^-$/s/pixel@25 $^{\circ}$C\\ 0.02 $\rm e^-$/s/pixel@-55 $^{\circ}$C\end{tabular} \\ \hline
        % Operation temperature& -50 $^{\circ}$C to 60 $^{\circ}$C   \\ \hline
        \end{tabular}
\end{table}

In this paper, the design of the focal plane detector is exhibited in section~\ref{CMOS camera design}. 
Next, the arrangement of the experimental is illustrated in section~\ref{experimental layout}. 
Then, the dark current and readout noise of the detector are studied in section~\ref{Readout_noise_dark_current}.
% After that, the split-pixel event characteristics, energy resolution, and other properties of the detector are investigated in section~\ref{data process}. 
After that, the split-pixel event characteristics, energy response of the detector, and mirror module alignment process with ``Burkert test'' method are investigated in section~\ref{data process}. 
Finally, we summarize the performance of the focal plane detector in section \ref{conclusion}.

\section{Design of X-ray focal plane detector}\label{CMOS camera design}
\begin{figure}[htb]
	\begin{center}
		\includegraphics[width=1\textwidth]{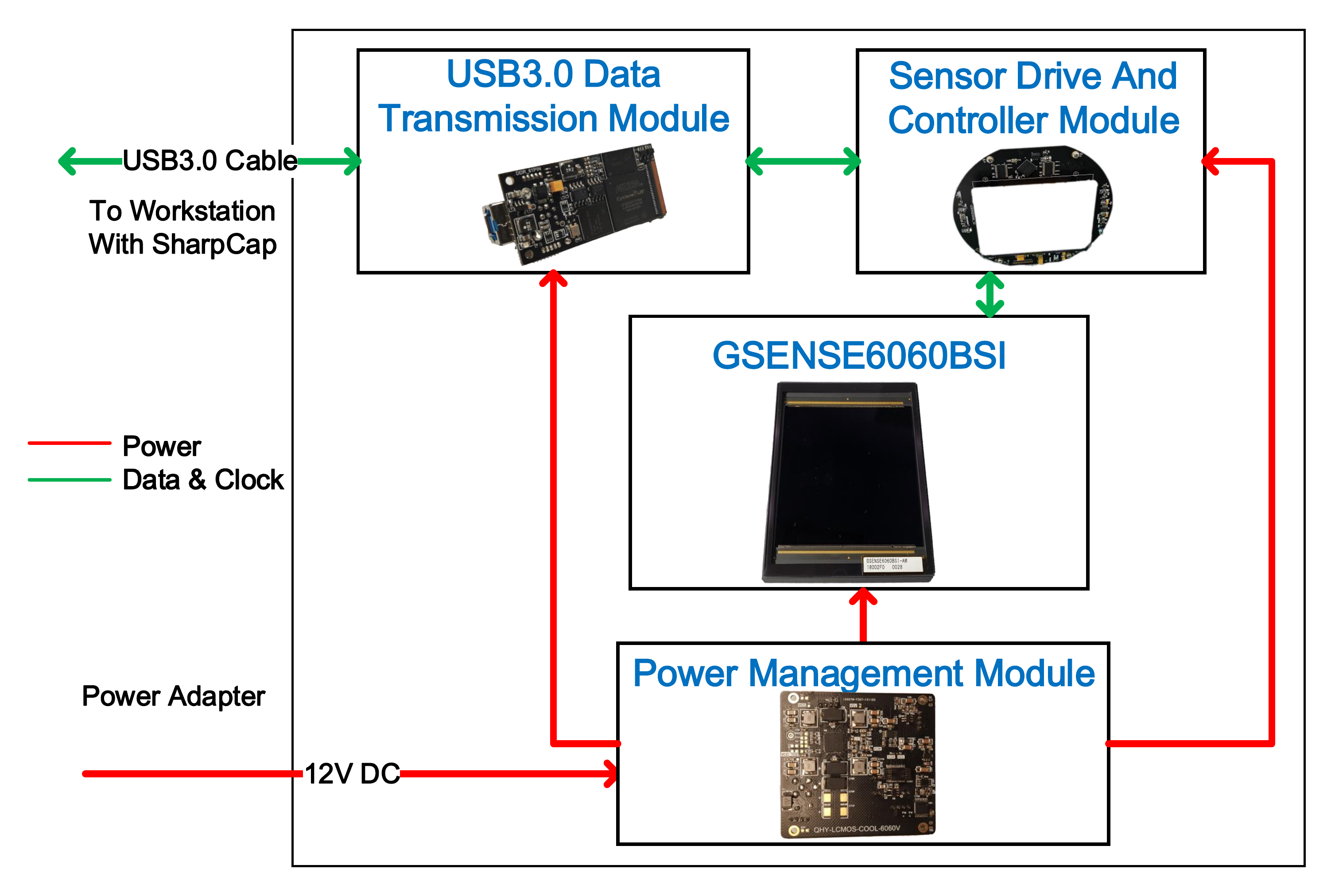}
		\caption{The block diagram shows the top-level design of the focal plane detector which mainly includes sCMOS sensor, sensor drive and controller module, USB3.0 data transmission module and power management module.}
		\label{6060_cricuit_block_diagram}
	\end{center}
\end{figure}

\begin{figure}[htb]
	\begin{center}
		\includegraphics[width=0.5\textwidth]{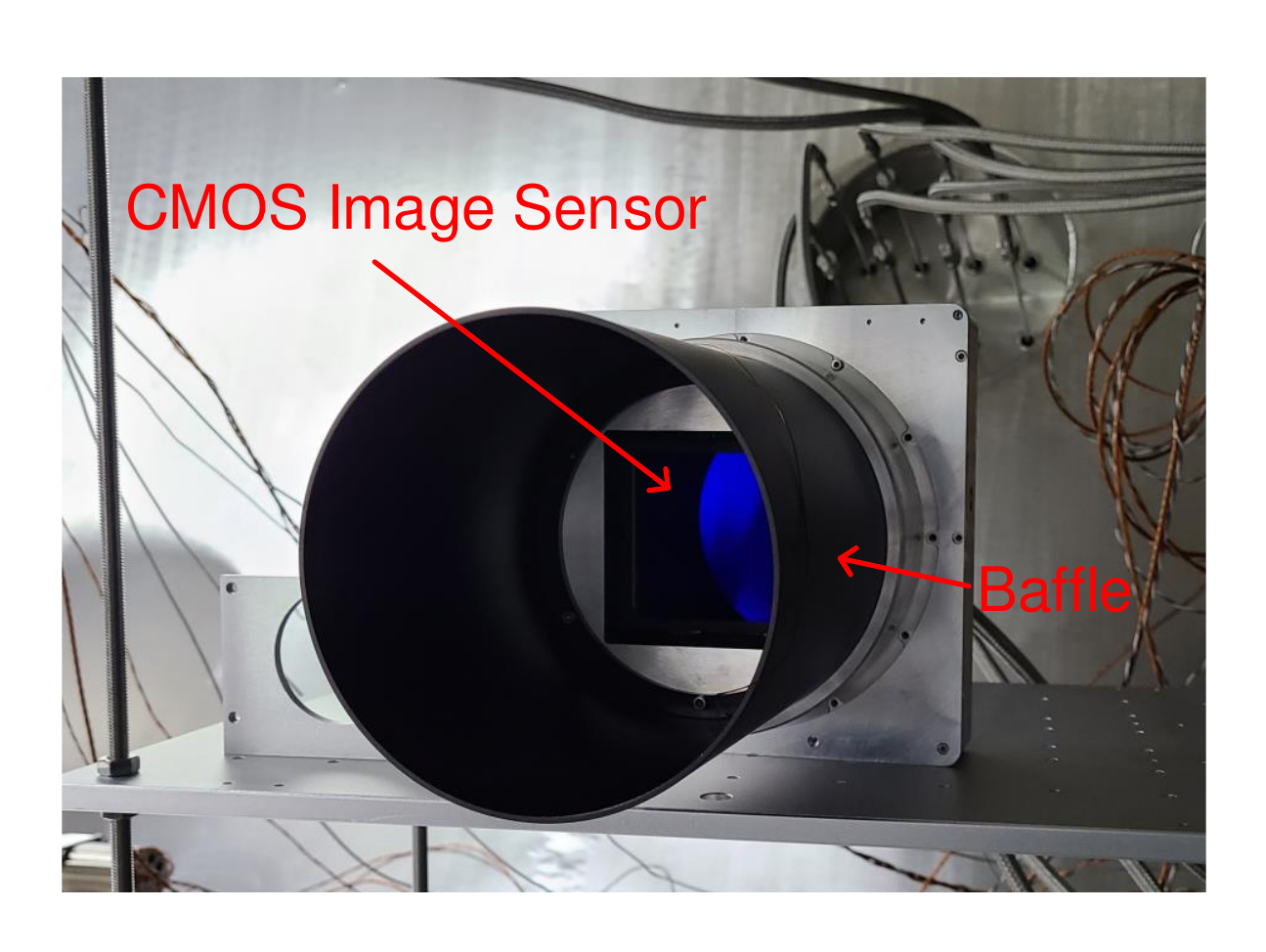}
		\caption{The picture of the focal plane detector. The black anodized aluminum alloy baffle is installed to reduce the influence of stray light.}
		\label{CMOS_camera_hood}
	\end{center}
\end{figure}

On the top-level design, the focal plane detector adopts a modular design, mainly composed of sCMOS sensor, sensor drive and controller module, data buffer and transmission module, power management module, as shown in Fig.~\ref{6060_cricuit_block_diagram}, which is convenient to upgrade and maintain in the future. 
% The focal plane detector should be moved to find the rough focus point during the alignment of the mirror module, thus the detector is wholly mounted on a precise translation stage in the vacuum chamber but not mounted on the flange \cite{Ling_2021} as usual. 
% In this case, regarding circuit design, Surface Mounted Components (SMD) are used to improve the heat dissipation efficiency in a vacuum. The Liquid electrolytic capacitors are forbidden to use in this circuits to avoid bursting caused by high vacuum (about $10^{-5}$ Pa). 
To reduce the X-ray absorption, the protection glass cover in front of the sensor is removed in advance in subsequent experiments. 
In addition, a 20 cm long black anodized aluminum alloy baffle is installed in front of the focal plane detector to reduce the influence of stray light, as shown in Fig.~\ref{CMOS_camera_hood}.

% SharpCap is a user-friendly and powerful astronomical camera capture tool software\cite{sharpcap}. 
SharpCap, an astronomical camera capture tool software \cite{sharpcap}, is used in this study to control and set the exposure parameters of the detector on the host computer through the USB3.0, and the detector is configured to work in ``14bit Standard'' mode \cite{GSENSE6060BSI_datasheet}.
Inside the detector, on receiving the exposure instruction from the host computer, the data transmission board will transfer the instruction to the drive and controller module to generate a specific drive timing. 
And then, the image data will be read out through the external data and clock LVDS interface of the sCMOS sensor. 
A piece of 512 Mbit DDR2 SDRAM (Double Data Rate 2 Synchronous Dynamic Random Access Memory) produced by \textit{Micron Technology, Inc.} is employed as a cache to store data temporarily to reduce the loss of image frames caused by data saturation, and then the data transmitted to the host computer through the USB3.0 cable.
The image could be displayed in real-time on the host computer, and the image data is stored as a FITS (Flexible Image Transport System) \cite{wells1979fits} standard format file.

\section{Experimental setup}\label{experimental layout}
\subsection{100 m long X-ray test facility}
% To calibrate the X-ray mirror module and other detectors of eXTP, 
A 100 m long X-ray test facility has been built in Beijing by the Institute of High Energy Physics, Chinese Academy of Sciences.
% , similar to PANTER \cite{BenediktMenz} of MPE (Max Planck Institute). 
The facility is mainly composed of an X-ray source station, a 100 m long vacuum tube, and a big vacuum chamber. 
The big vacuum chamber with a cylindrical shape is placed at the end side of the 100 m vacuum tube. 
The inner diameter and length of the big vacuum chamber are about 3.4 m and 8 m, respectively \cite{zhao2019Wolter}. 
The big vacuum chamber is equipped with multiple mechanical pumps, molecular pumps, and cryogenic pumps, which make the vacuum capable of reaching the order of $10^{-5}$ Pa. 
Liquid nitrogen or cold nitrogen gas is equipped to cool the detector as low as -100$^{\circ}$C with a jitter of less than 1$^{\circ}$C. 

\subsection{Multi-target fluorescence X-ray source}
\begin{figure}[htb]
	\begin{center}
        \includegraphics[width=0.8\textwidth]{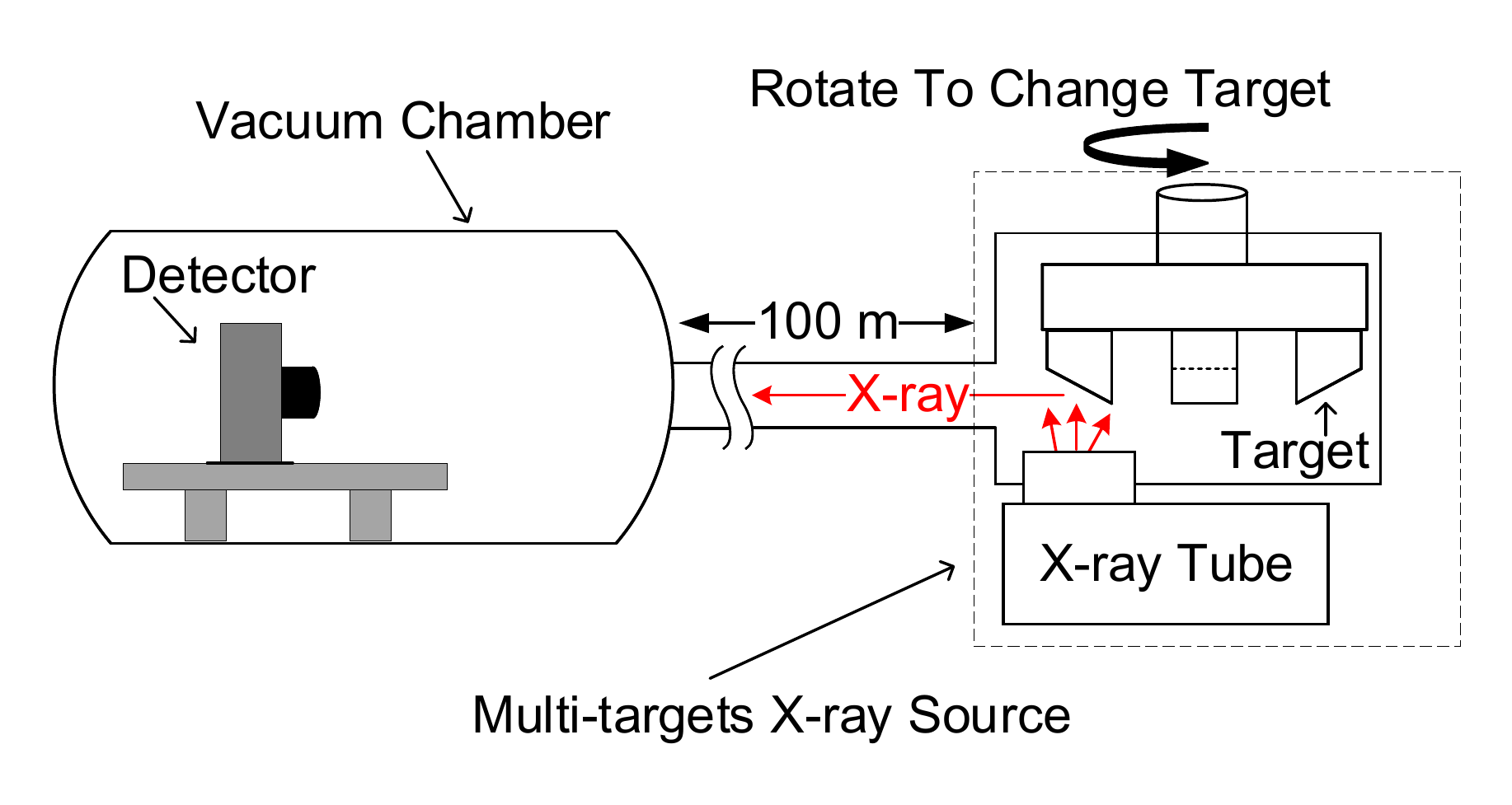}
		\caption{The experimental layout block diagram. The focal plane detector is placed in the big vacuum chamber. The multi-target fluorescence X-ray source can be rotated to different targets to obtain various characteristic lines. (Note: The vacuum chamber and the multi-target fluorescence X-ray source in the block diagram are not plotted as the actual ratio.)}
		\label{experimenta_setup}
	\end{center}
\end{figure}

The starting end of the 100 m vacuum tube is equipped with a self-developed multi-target fluorescence X-ray source (including C, $\rm SiO_2$, Al, Mg, Mo, Ti, Cr, Fe, Cu, covering the energy from 0.2 to 10 keV). 
The multi-target fluorescence X-ray source is made by putting a high-purity target in front of the output window of a traditional X-ray tube.
The X-rays (including characteristic lines and bremsstrahlung X-ray spectroscopy) generated by the traditional X-ray source are used to irradiate and excite the secondary target, and only the characteristic line of the target material is generated during the de-excitation process in the secondary target. 
The experimental layout is shown in Fig.~\ref{experimenta_setup}.

\section{Readout noise and dark current}\label{Readout_noise_dark_current}
Readout noise and dark current are mainly influenced by exposure time and temperature, which affect the energy resolution and the threshold of the lowest detection energy.
However, the photon statistics on a frame decrease with the decreasing of the exposure time. So, appropriate working conditions should be set.

\subsection{Readout noise}
\begin{figure}[htb]
	\begin{center}
		\includegraphics[width=0.6\textwidth]{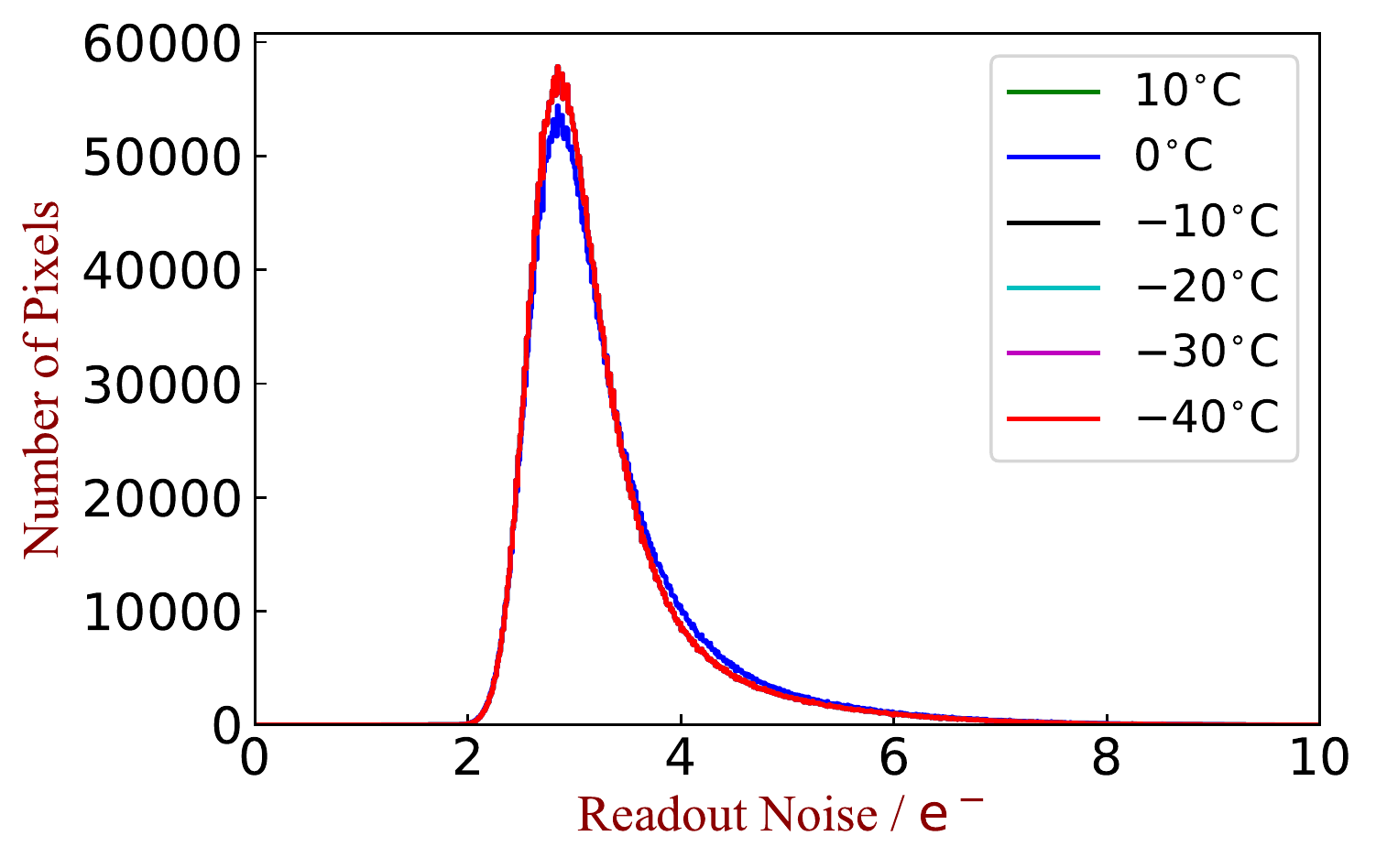}
		\includegraphics[width=0.6\textwidth]{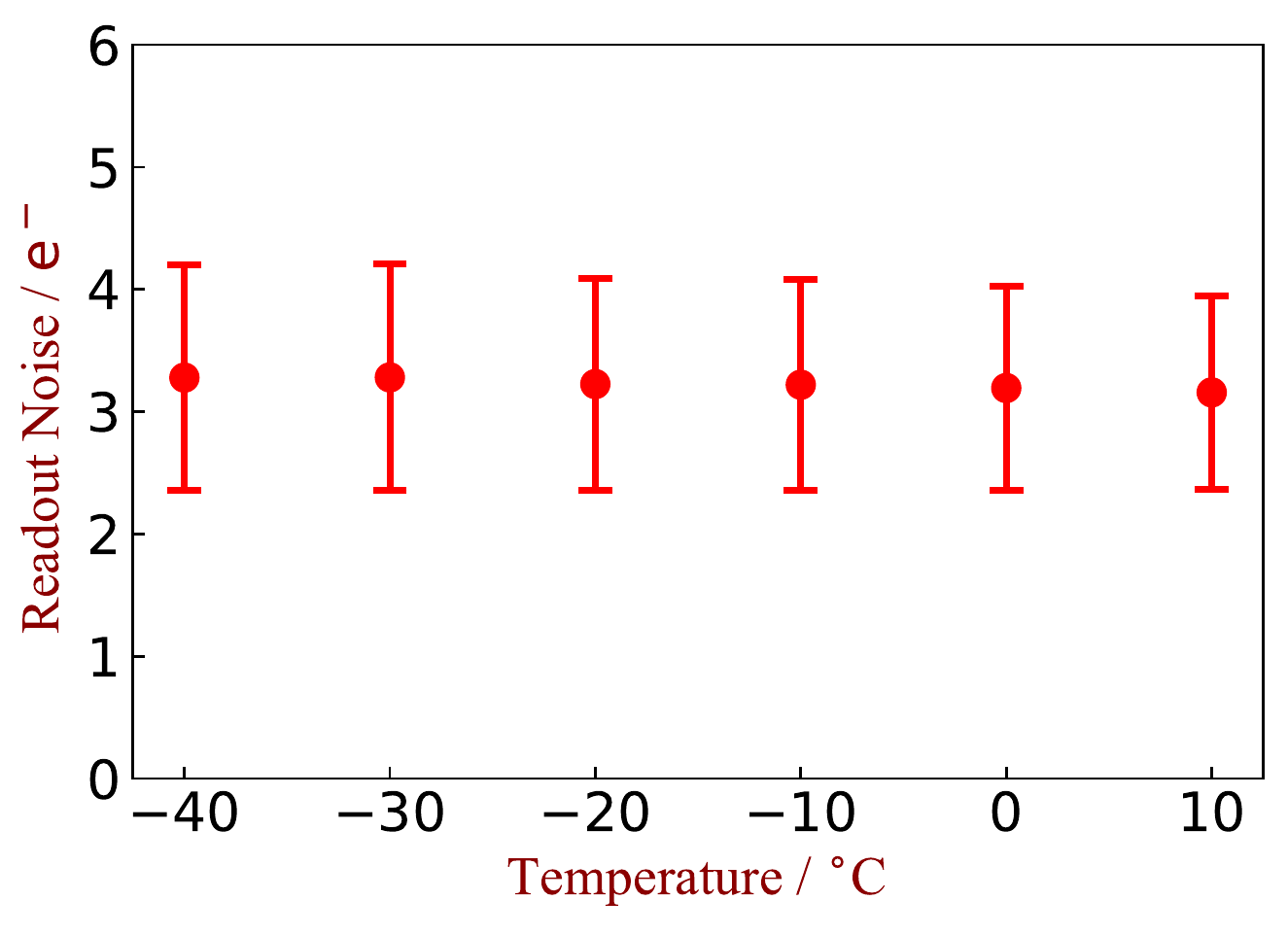}

		\caption{
		Top panel, the readout noise distribution of each pixel at different temperatures. 
		Bottom panel, The readout noise of the detector is about 3.2 $\rm e^-$ and the RMS is about 0.8 $\rm e^-$ according to Eq.~\ref{electron gain}, the RMS of $\sigma_i$ is treated as the error bar of readout noise. 
		The readout noise has a slight increase at low temperatures, which may be related to the interior structure of the pixels \cite{Wang_2019}.}
		\label{readout_noise}
	\end{center}
\end{figure}

Readout noise is one of the main noise sources in sCMOS sensors, and it affects the energy detection limit  of the detector.
To reduce the integral of dark current, the exposure time is set as short as possible to 1 ms, and 100 frames are acquired in different temperatures. 
The root mean square (RMS) of the value of each pixel in a selected area (seen in section~\ref{single/split pixel event}), with an unit of Analog Digital Unit (ADU), is calculated to get $\sigma_i$ ($i$ takes from 1 to N, N equals the number of total pixels). The distribution of $\sigma_i$ is shown in Fig.~\ref{readout_noise} (Top panel). The median of $\sigma_i$ is usually reported as the readout noise of the sensor \cite{LiLuchang2016}, as shown in Fig.~\ref{readout_noise} (Bottom panel), and the readout noise charge is about 3.2 $\rm e^-$ according to eq.~\ref{electron gain}. 

\subsection{Dark current}
\begin{figure}[htb]
	\begin{center}
		\includegraphics[width=0.6\textwidth]{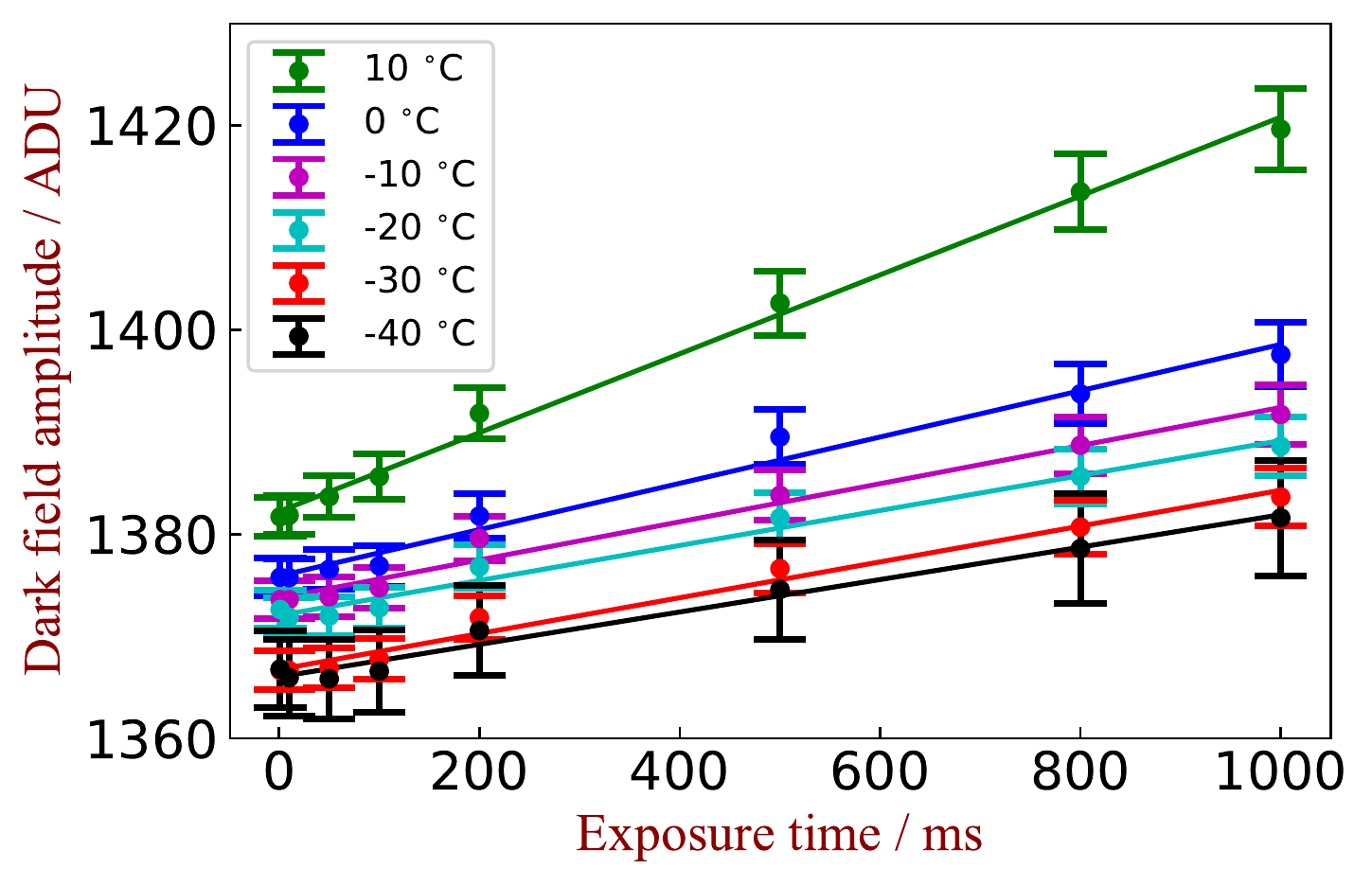}
		\caption{The dark field amplitude obtained by Gaussian fitting the pixel value distribution of all pixels in the selected area (seen in section~\ref{single/split pixel event}) as function of exposure time.
		The dark current of the detector is the slope of a line fitted at different exposure times with the same temperature. 
		The length of the error bar equals the $\sigma$ of Gaussian fitting. 
		The dark current is nearly unchanged when the temperature is below -30 $^{\circ}$C and with the dark current of about 18 $\rm e^-/pixel/s$ according to Eq.~\ref{electron gain}.}
		\label{Dark_current}
	\end{center}
\end{figure}

To improve the performance of energy resolution and energy detection limit, the thermal noise should be reduced as possible \cite{OGINO2021164843}\cite{9060018}. Hence, The dark current properties of the detector is studied. 
The exposure time is set to 1 ms, 10 ms, 50 ms, 100 ms, 200 ms, 500 ms, 800 ms, and 1000 ms, respectively, and 100 dark frames are acquired in each temperature. 
The dark field amplitude is the mean obtained by Gaussian fitting the pixel value distribution of all pixels in a selected area (seen in section~\ref{single/split pixel event}).
The dark current is the slope of a line fit at different exposure times with the same temperature, as shown in Fig.~\ref{Dark_current}. 
The dark current is nearly unchanged when the temperature is below -30 $^{\circ}$C, maintaining a dark current of about 18 $\rm e^-/pixel/s$. 
The dark current is slightly larger than expected in Table~\ref{tab_specification_6060}, which may be caused by the temperature measured being the backside (i.e. cold side) of the sCMOS sensor instead of the sensor surface.

\section{Results}\label{data process}
\subsection{Single-pixel event and split-pixel event}\label{single/split pixel event}
\begin{figure}[htb]
	\begin{center}
		\includegraphics[width=0.6\textwidth]{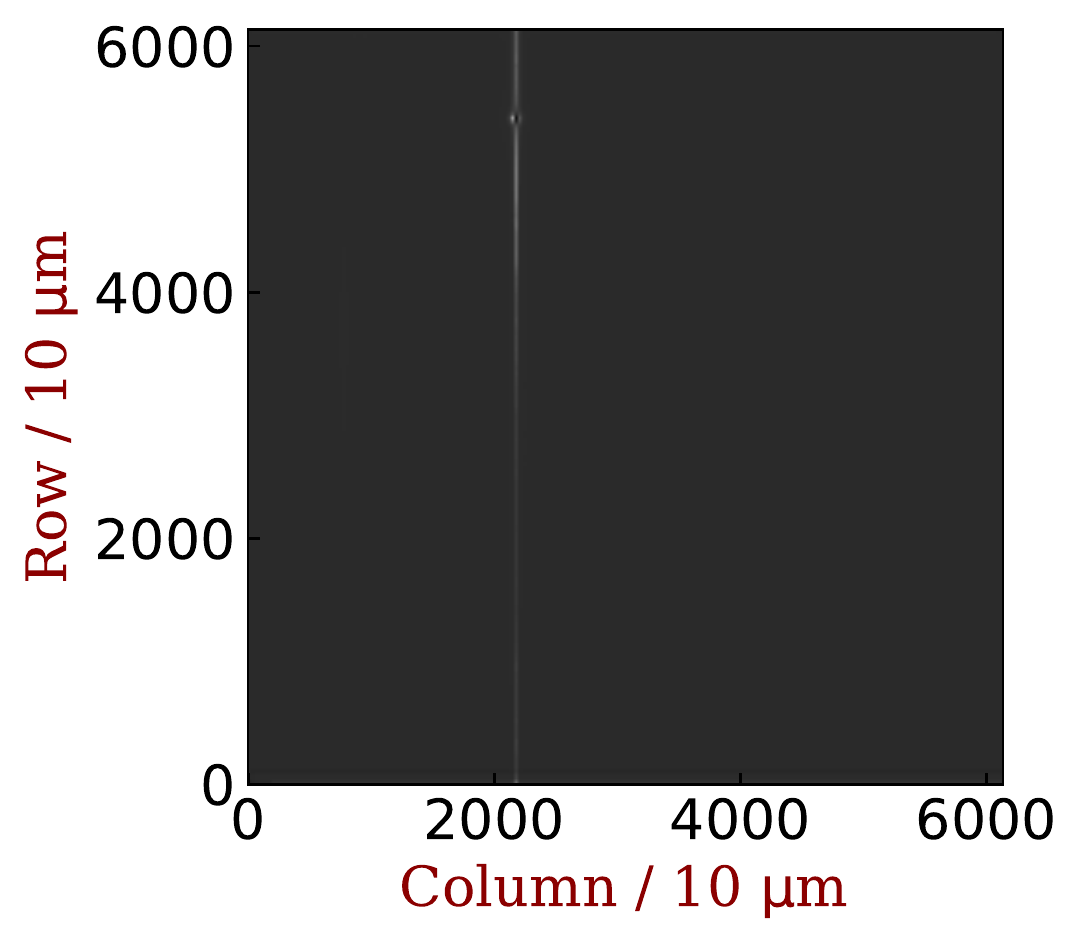}
		\caption{One frame of dark field image is acquared by setting the exposure time to 100 ms with a temperature of -30 $^{\circ}$C. An obvious bad line near the 2000th column. The blemish luminescence generated by the bad line becomes more obvious as the exposure time increases.}
		\label{raw_dark_image}
	\end{center}
\end{figure}

\begin{figure}[htb]
	\begin{center}
		\includegraphics[width=0.6\textwidth]{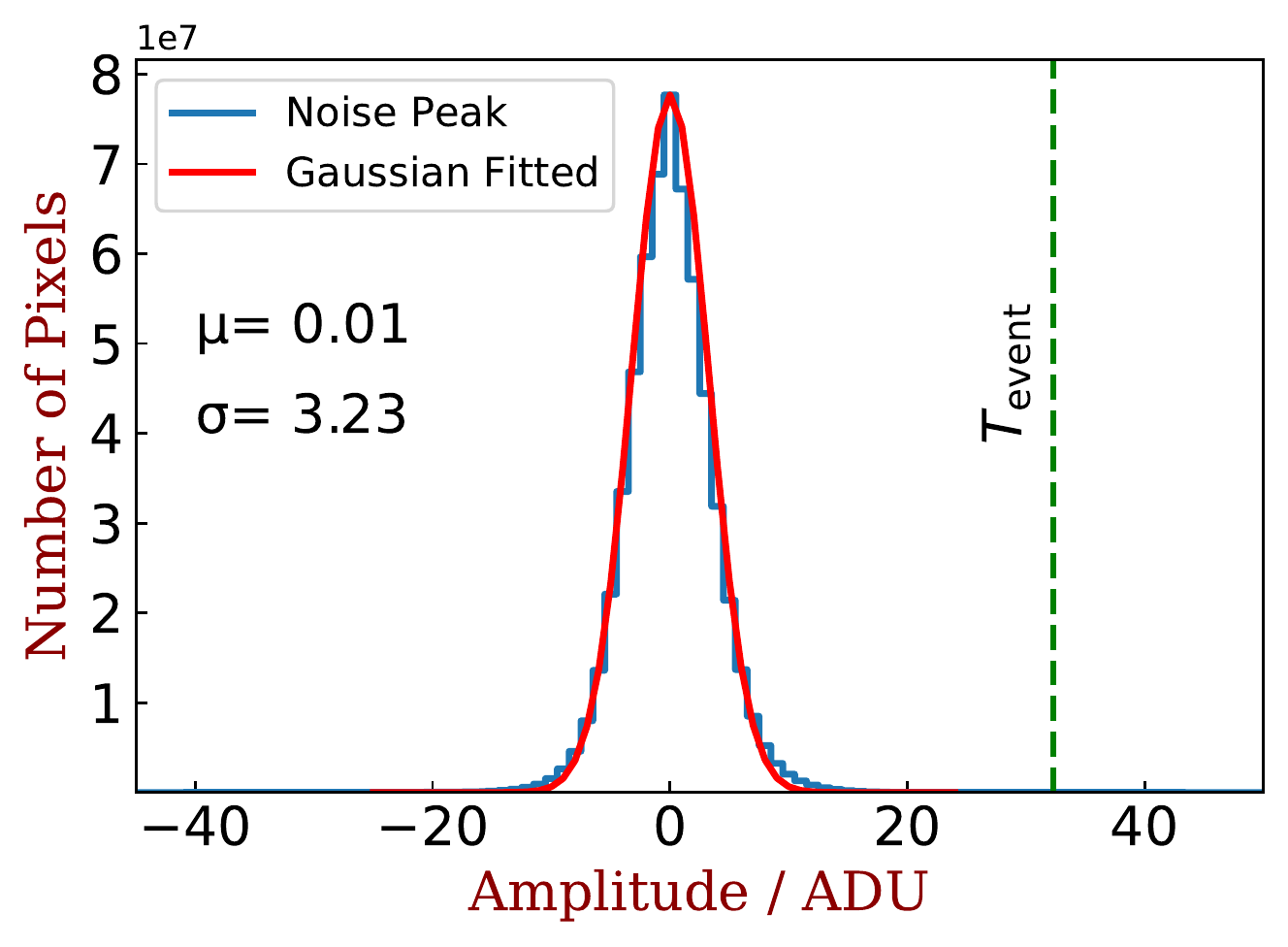}
		\caption{The noise distribution of the detector in dark field subtracted by bias map $I_{\rm bias}$. 
		The blue line represents the noise peak obtained by setting the exposure time to 100 ms with a temperature of -30 $^{\circ}$C. 
		The red line represents the result of Gaussian Fitting, and we get the $\upmu$=0.01, $\upsigma$=3.23. 
		The threshold of X-ray photon events and noise signal discrimination is set to 10$\times\sigma$ of the noise peak ($T_{\rm event}$), because the noise is mainly below 30 ADU. 
		According to the E-C relation in Eq.\ref{E-C relation}, the low detect limitation equals 10$\times\upsigma\times{R_{\rm E-C}}$+74.31= 186 eV (indicated with a green dash line). 
% 		The Root-Mean-Square (RMS) of noise is 3.229$\times$(1.04$\rm e^-$/Channel)=3.4$\rm e^-$/pixel.
		}
		\label{biased_noise_peak}
	\end{center}
\end{figure}

% multi-target fluorescence X-ray source (including Al, Mg, Ti, Fe, Cu targets) was adopted to study the characteristic of split pixel event of GSENSE6060BSI. 
% An X-ray photon interacts with the photodiode (PD) inside a pixel to generate electron-hole pairs which drift to the collector under the action of an electric field, and then subsequent signal amplification and digitization processing will be performed. 
Considering the limited strength of the electric field in the photodiode of the sCMOS sensor, there is a significant diffusion effect during the drift process of the electron-hole pair generated by X-ray.
Thus, the electron-hole pairs will be shared by two or more adjacent pixels when the interaction position of the X-rays is close to the edge of a pixel, resulting in an incomplete charge collection (ICC) effect in a pixel\cite{OGINO2021164843}, and may even misjudge one photon event as several photon events . 
The process of split-pixel event reconstruction is as follows:

\begin{enumerate}
    \item 
    % The grade of GSENSE6060BSI sensor we used in this study is ``grad 1"
    Only the part of the area far away from the bad lines (those pixels in the same readout column cannot work properly), as shown in Fig.~\ref{raw_dark_image}, is selected for the energy spectrum processing (i.e. the area of 3000th to 5000th columns and 1000th to 4000th rows is selected) to avoid the influence of blemish luminescence \cite{JamesCCD}. 
    In addition, when the sensor is in a dark field, the value of pixels larger than 20 times $\sigma$ of the noise peak is regarded as bad pixels or noisy pixels and will be discarded. 
    
    \item 
    Both the non-uniformity of the pixel during manufacturing and the difference of Analog Digital Converter (ADC) and Programmable Gain Amplifier (PGA) of each channel in rolling shutter structure contribute to the different response of pixels to photons \cite{WILLEMIN2001185}\cite{hasegawa2009development}. 
    To reduce the effect of non-uniformity, the exposure time is set to 100 ms under the temperature of -30 $^{\circ}$C, and obtains 100 frames when the sCMOS sensor is in darkness. 
    Calculate the median of pixel value with the same position in 100 frames, and get the bias map $I_{\rm bias}$ which presents the inconsistency between different pixels \cite{pnccd}:
    \begin{equation}
    % {bmatrix}
        I_{\rm bias}=
        \begin{bNiceMatrix}
            m_{1,1} & \Cdots& m_{1,\rm{j}} \\
            \Vdots & \Ddots & \Vdots \\
            m_{\rm{i},1} & \Cdots & m_{\rm{i,j}} 
        \end{bNiceMatrix}        
    \end{equation}
    where, $m$ is the median of a corresponding pixel, i and j equals 2000 and 3000, respectively.

    \item 
    Before evaluating the split-pixel event, it is necessary to distinguish the X-ray photon events from the noise signals. 
    100 dark field images are subtracted by bias map $I_{\rm bias}$ to remove the influence of dark current. 
    Then, the noise peak is fitted with a Gaussian function, as shown in Fig.~\ref{biased_noise_peak}. 
    Set the threshold of X-ray photon events and noise signal discrimination $T_{\rm event}$ to 10 times $\sigma$ of the noise peak. Pixel value will be considered as a contribution from X-ray photon events when the pixel value is greater than $T_{\rm event}$, otherwise will be treated as noise (i.e. the threshold is set to 186 eV according to Eq.~\ref{E-C relation}). 
    
    \item 100 frames of images with characteristic lines of Mg, Al, Ti, Fe, and Cu are acquired, respectively. 
    Images with X-ray photon events are subtracted by bias map $I_{\rm bias}$, and then judge the X-ray photon events from noise according to $T_{\rm event}$. 
    An event whose value exceeds the event threshold $T_{\rm event}$ while the value of its adjacent pixels (excluding diagonal neighbors) are below the threshold $T_{\rm event}$ will be considered as single-pixel event. 
    On the contrary, an event will be judged as an split-pixel event if the value of the pixel and its surrounding 8 pixels are grater than $T_{\rm event}$.

    \item  
    The module dedicated to numerical calculation and analysis in Python, Scipy, has the ``ndimage'' class \cite{scipy} for multi-dimensional image processing. By calling the ``binary$\_$hit$\_$or$\_$miss'' function, the position of a given pattern will be returned. 
    Thus, single-pixel event selection and split-pixel event reconstruction are available as long as the patterns of the split-pixel event can be enumerated.
    Since the proportion of generated charge distributes into more than 4 pixels is less than 1$\%$, the split-pixel event is divided into three categories: 2-pixel split event, 3-pixel split event and 4-pixel split event.
    
    \item  
    Once the coordinates of the single-pixel event or $n$-pixel split event ($n$ equals 2, 3, or 4) are given, sum up the maximum pixel value and the values of the surrounding 8 pixels to reconstruct the incident X-ray photon events.
    
\end{enumerate}

\begin{figure}[htb]
	\begin{center}
		\includegraphics[width=0.8\textwidth]{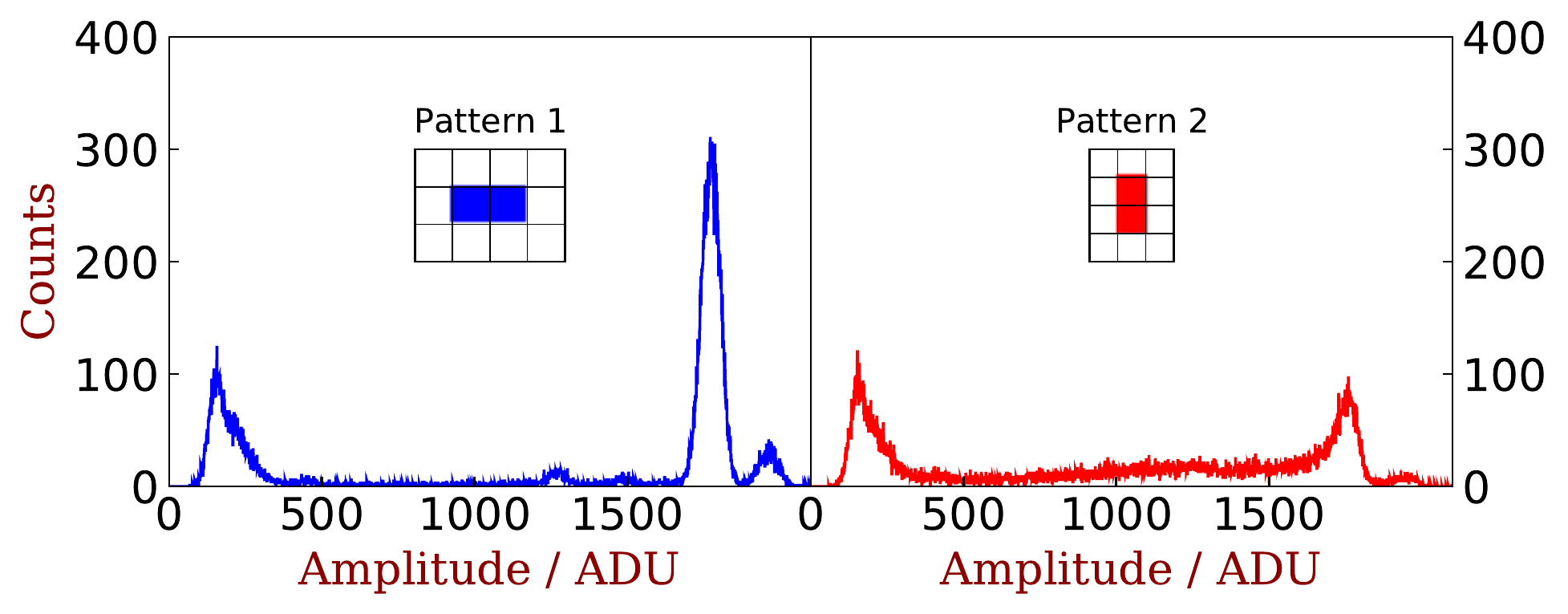}
		\caption{By reconstructing the incident X-ray photons of the Fe characteristic line with 100 frame image data, the pixel value distribution of 2 possible patterns in the 2-pixel split event is obtained.}
		\label{double_evnet(Fe)}
	\end{center}
\end{figure}

\begin{figure}[htb]
	\begin{center}
		\includegraphics[width=1\textwidth]{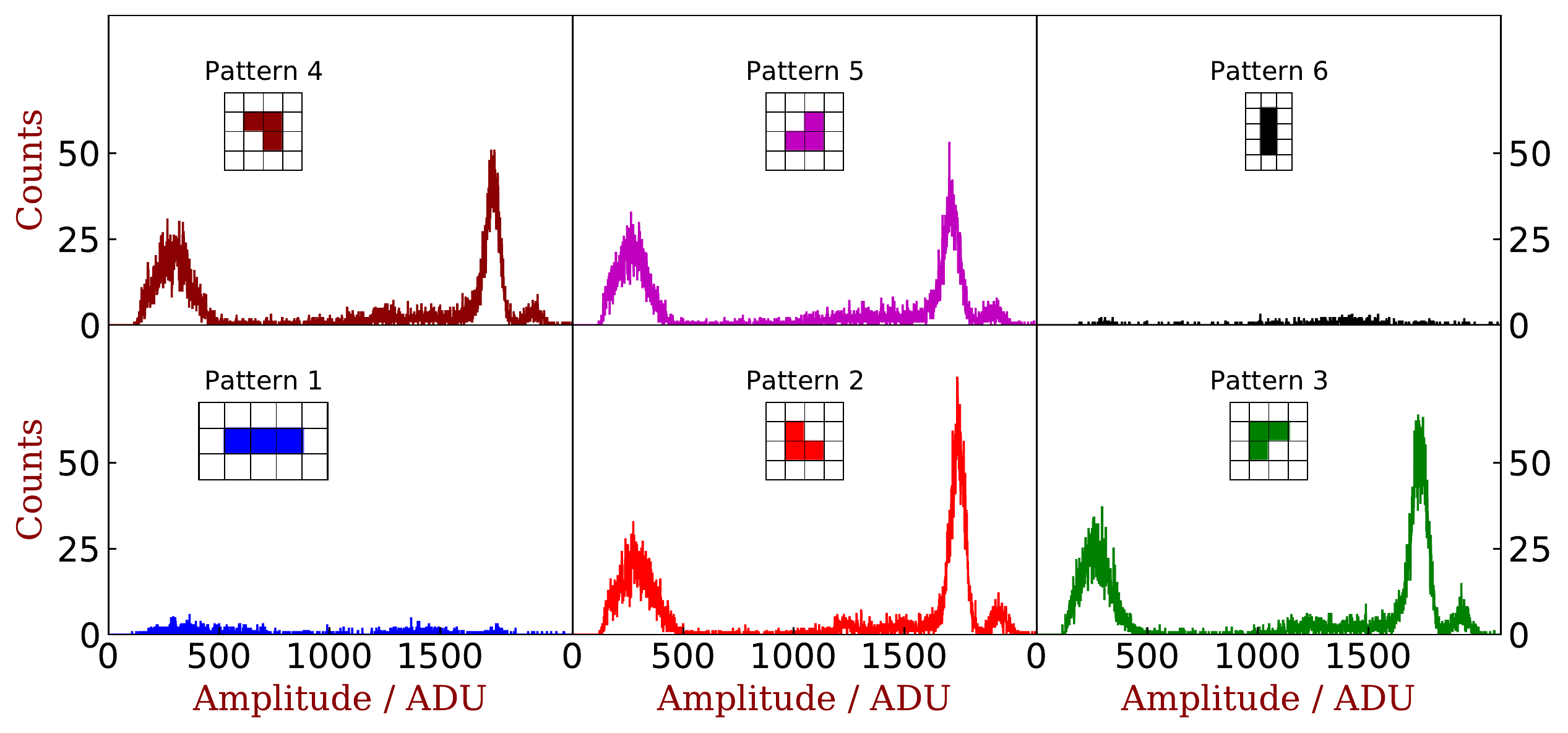}
		\caption{By reconstructing the incident X-ray photons of the Fe characteristic line with 100 frame image data, the pixel value distribution of 6 possible patterns in the 3-pixel split event is obtained.}
		\label{triple_evnet(Fe)}
	\end{center}
\end{figure}

\begin{figure}[htb]
	\begin{center}
    	\includegraphics[width=1\textwidth]{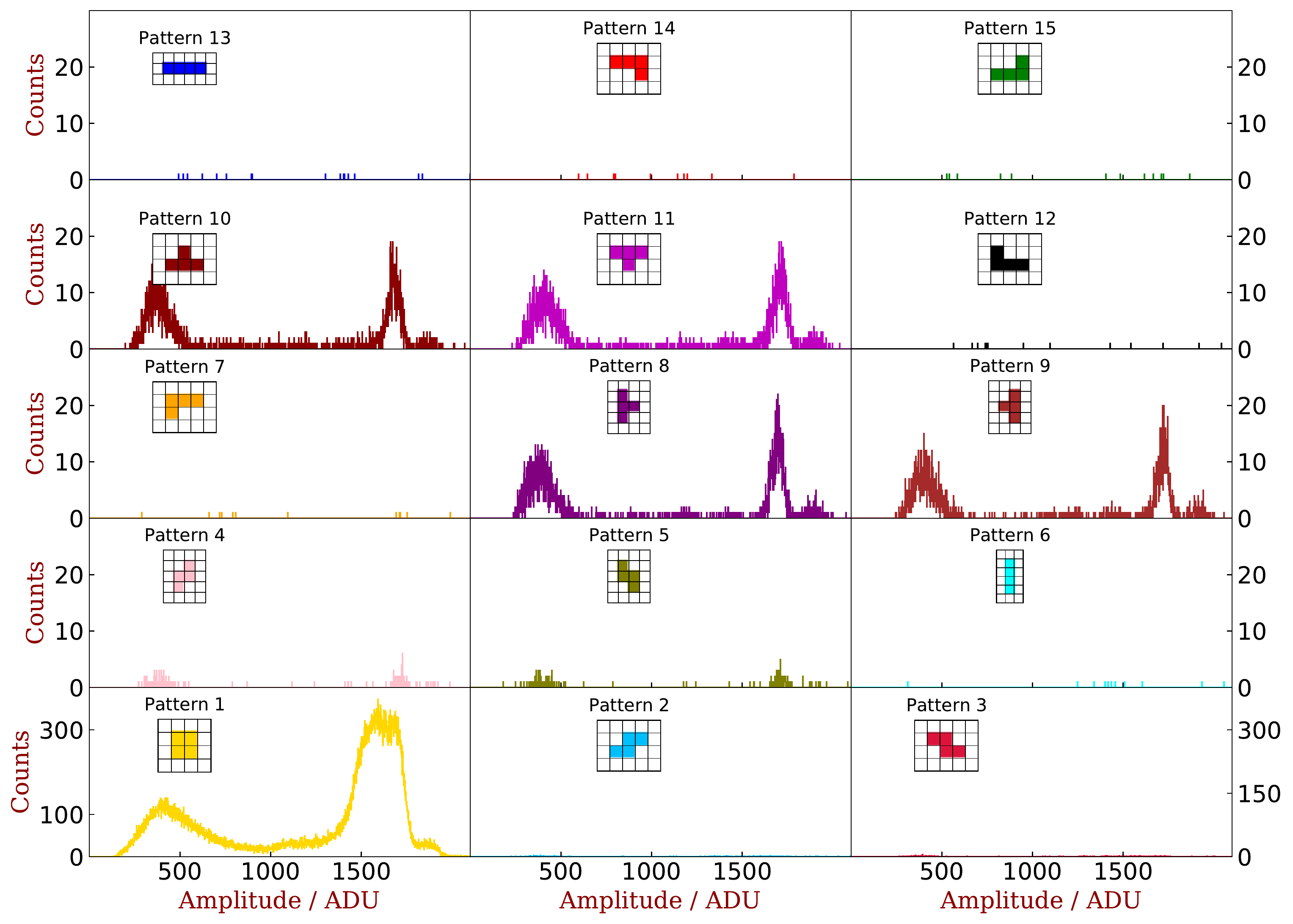}
	\caption{By reconstructing the incident X-ray photons of the Fe characteristic line with 100 frames image data, the pixel value distribution of 15 possible patterns in the 4-pixel split event is obtained.}
	\label{quadruple_evnet(Fe)}
	\end{center}
\end{figure}

Taking the reconstruction result of the Fe characteristic line as an example, there are 2, 6, 15 possible patterns for 2-pixel split event, 3-pixel split event, 4-pixel split event, respectively, as shown in Fig.~\ref{double_evnet(Fe)}, Fig.~\ref{triple_evnet(Fe)}, and Fig.~\ref{quadruple_evnet(Fe)}. 
As we can see, there is nearly no photon in some possible patterns, thus, those reconstructed events will not be taken into account. Especially, considering the event reconstruction of ``pattern 1'' in 4-pixel split event (see in Fig.~\ref{quadruple_evnet(Fe)}) is deteriorated, those counts belonging to ``pattern 1'' are discarded even though it occupies about 92.7$\%$ of all 4-pixel split events. 

\begin{figure}[htb]
	\begin{center}
		\includegraphics[width=0.8\textwidth]{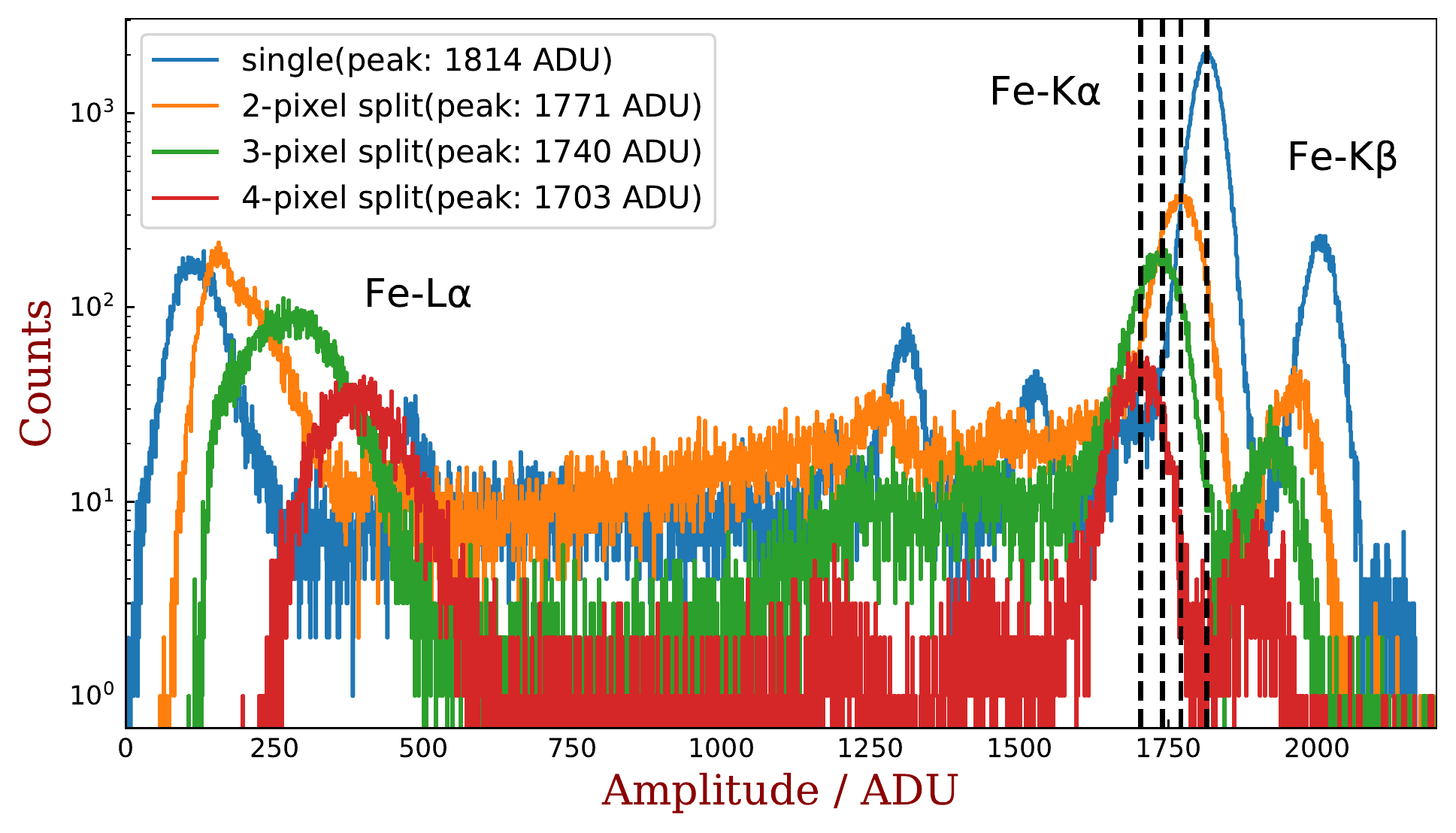}
		\caption{The pixel value distribution of single-pixel events and split-pixel events under the irradiation by Fe characteristic line. 
		For 2-pixel split event, pattern 1 and 2 are superimposed shown in Fig~\ref{double_evnet(Fe)}. 
		For 3-pixel split event, pattern 2, 3, 4 and 5 are superimposed shown in Fig.~\ref{triple_evnet(Fe)}. 
		For 4-pixel split event, pattern 8, 9, 10 and 11 are superimposed shown in Fig.~\ref{quadruple_evnet(Fe)}.
		The more split of generated charge to adjacent pixels, the worse the energy resolution.
		The counts and proportions of the split-pixel events are shown in Table~\ref{tab:split_event_percentage}. 
		The peak of Fe-K$\upalpha$ is indicated with black dash line and its value is shown in the legend.
		}
		\label{split_event(Fe)}
	\end{center}
\end{figure}

Through event reconstruction, the characteristic lines of Fe-$\rm L{\upalpha}$, Fe-$\rm K{\upalpha}$, Fe-$\rm K{\upbeta}$ are shown in Fig.~\ref{split_event(Fe)}. 
First, due to the ICC during its spread to adjacent pixels \cite{HARO2020108354}, the peak position of Fe-$\rm K{\upalpha}$ is decreased. The Fe-$\rm K{\upalpha}$ peak position of single-pixel event, 2-pixel split event, 3-pixel split event, and 4-pixel split event is 1814 ADU, 1771 ADU, 1740 ADU, and 1703 ADU, respectively. 
Then, the process of selecting an $n$-pixel split event raises the threshold of a reconstructed event (i.e. the minimum reconstructed photon energy of the single-pixel event, 2-pixel split event, 3-pixel split event, and 4-pixel split event are $T_{\rm event}$, 2 $\times T_{\rm event}$, 3$\times T_{\rm event}$, and 4$\times T_{\rm event}$, respectively).
The proportions of the different event split patterns show significant differences in each other, which maybe representing the polarization properties of incident X-ray photons according to the research by Kazunori Asakura et al. 2019 \cite{asakura2019x}.

\subsection{Energy calibration of detector}
\begin{figure}[htb]
	\begin{center}
		\includegraphics[width=0.8\textwidth]{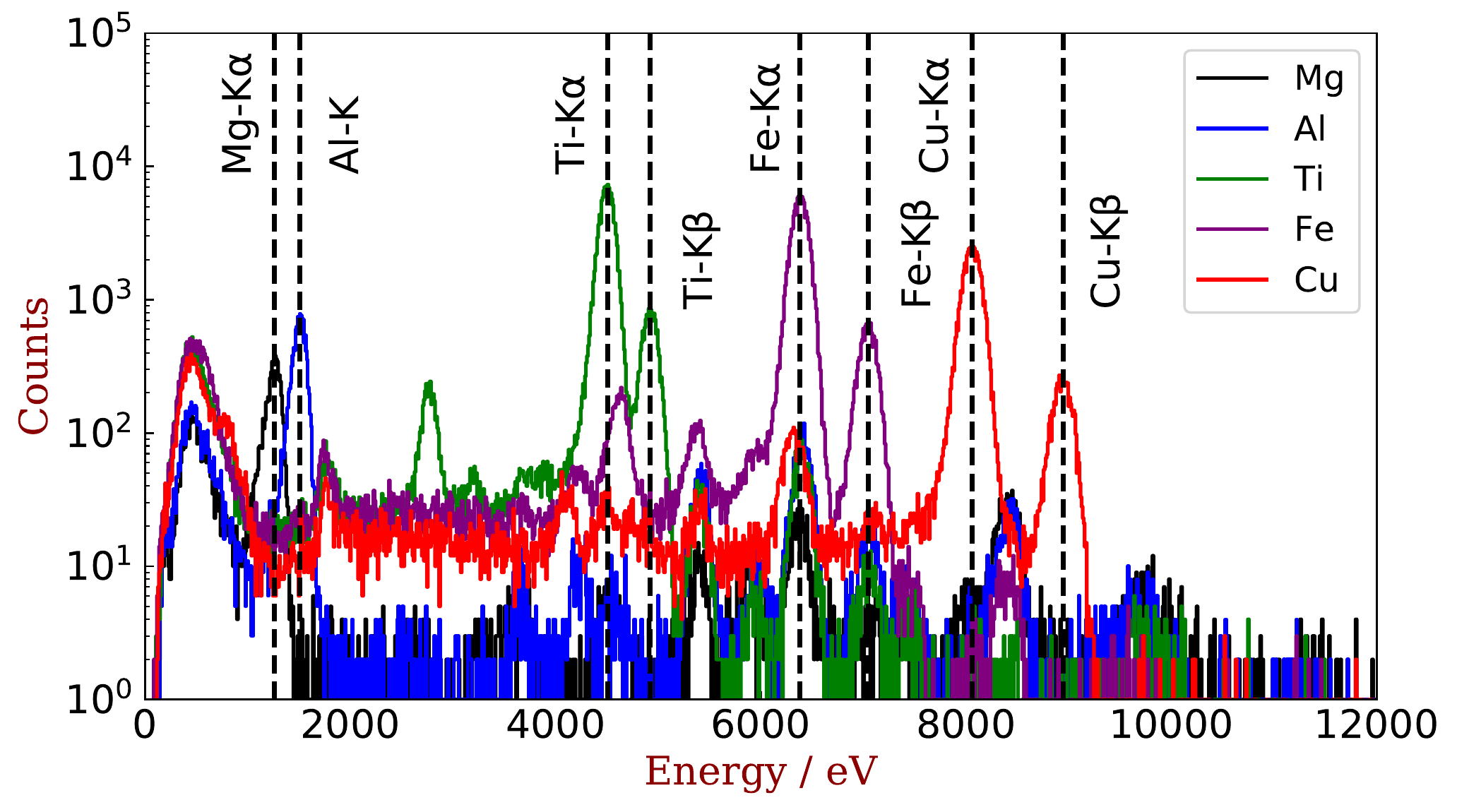}
		\includegraphics[width=0.8\textwidth]{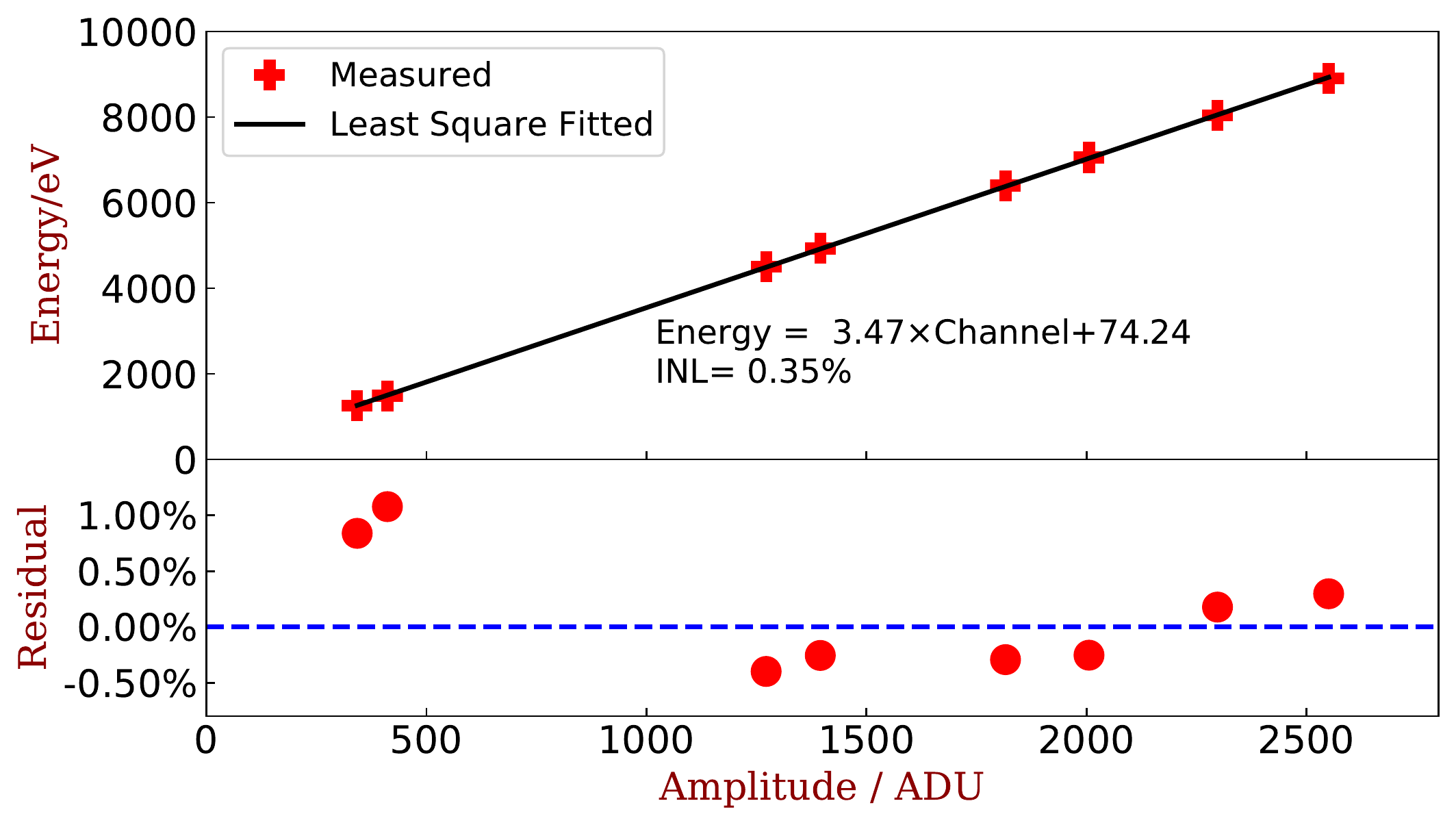}
		\caption{
		Top panel: The energy distribution of single-pixel events are tested with the characteristic lines of Mg, Al, Ti, Fe and Cu at a temperature of -30 $^{\circ}$C, which cover the working energy range of the eXTP mirror module.
		Bottom panel: The relationship between energy and amplitude is obtained by least square fitting, and the integral nonlinearity (INL) is about 0.35$\%$.}
		\label{sigle_event_spectrum}
	\end{center}
\end{figure}

Taking the ICC effect during charges spread to adjacent pixels into account, only single-pixel events are selected to determine the energy to channel (E-C) relationship. 
The characteristic lines of Al, Mg, Ti, Fe and Cu are shown in Fig.~\ref{sigle_event_spectrum} (Top panel), and the E-C relationship of focal plane detector is calibrated:
\begin{equation}
    E = 3.47 \times C + 74.24
    \label{E-C relation}
\end{equation}
where, $E$ is deposition energy (with an unit of eV), $C$  is the mean of a characteristic line fitted with Gaussian function (with an unit of ADU), as shown in Fig.~\ref{sigle_event_spectrum} (Bottom panel). 
The fitted line which does not pass through the zero point may result from the non-uniformity of the sCMOS sensor.

\begin{figure}[htb]
	\begin{center}
		\includegraphics[width=0.6\textwidth]{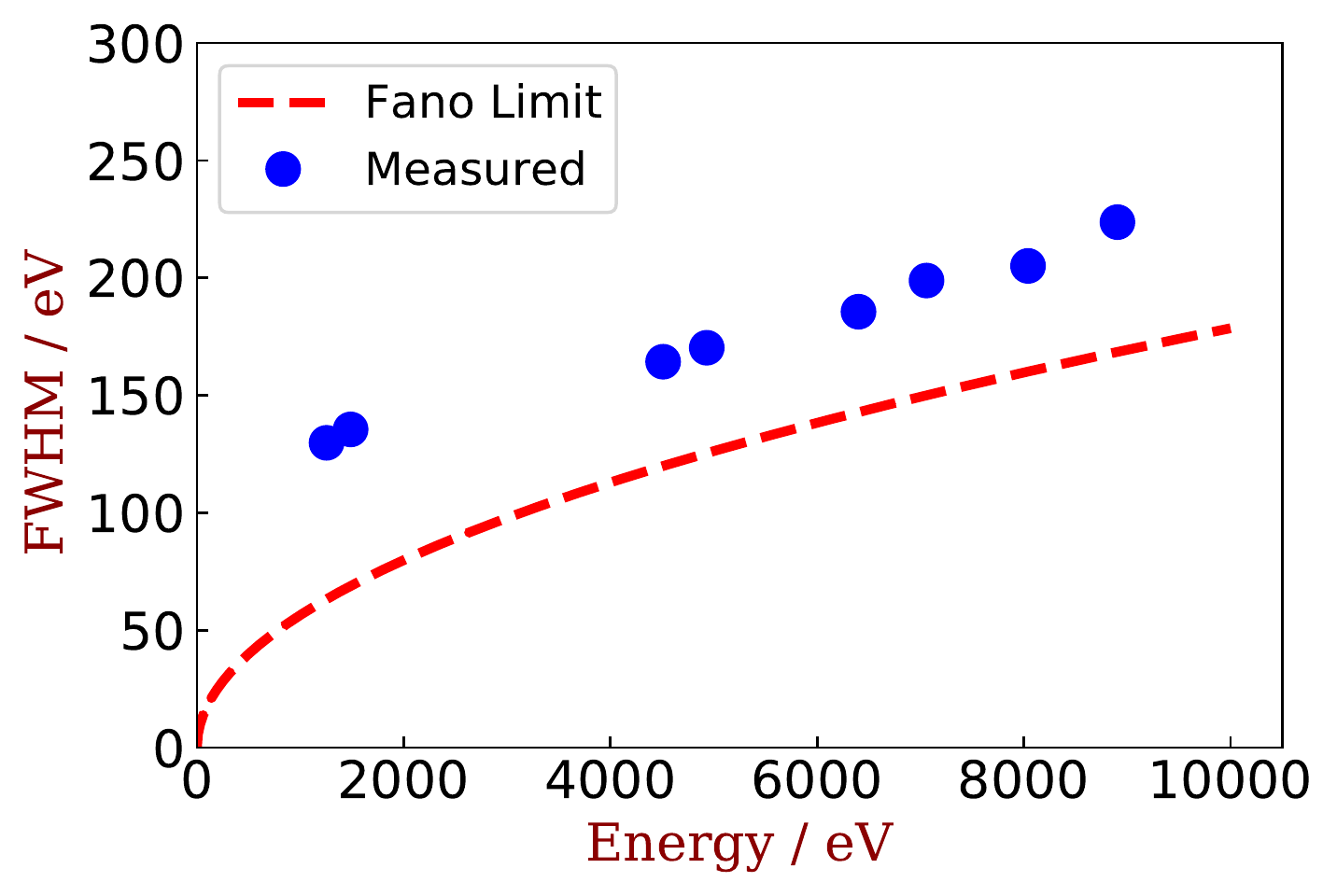}
		\caption{
		The energy resolution of the detector for the single-pixel events.
		The blue dot represents the measured energy resolution of the detector.
		The red dash line represents the Fano limit of the Si-based detector.
		The FWHM reaches 199 eV@8.04 keV (Cu-K). 
		The error of energy resolution of the focal plane detector is not given for it is small.
		}
		\label{FWHM_vary_with_energy}
	\end{center}
\end{figure}

The gain of electron can be calculated: 
\begin{equation}
    G = W / {R_{\rm E-C}}
    \label{electron gain}
\end{equation}
where, $W$ is the average ionization energy. $W$ equals 3.71 eV/$\rm e^-$ at temperature of -30 $^{\circ}$C \cite{PEHL196845}. $R \rm{_{E-C}}$ is the E-C conversion factor in Eq.~\ref{E-C relation}. $R \rm{_{E-C}}$ equals 3.47 eV/ADU. Finally, the gain of electron equals 1.07 $\rm e^-$/ADU. 
Thus, the energy resolution of the focal plane detector can be calculated, as shown in Fig.~\ref{FWHM_vary_with_energy}. 
The measured energy resolution is worse than the Fano limit \cite{PhysRev.72.26}\cite{OWENS2002437}, which may be caused by the non-uniformity of pixels and channels gain. \cite{NARUKAGE2020162974}. 
Thus, the energy resolution can be improved by the gain correction of the different readout channels in the future.

\begin{table}[htb]
    \centering
    \caption{Counts and its proportions of single pixle event and $n$-pixel split event}
    \label{tab:split_event_percentage}
        \begin{tabular}{ccc}
        \hline
        Event type & Counts & Proportion \\ \hline
        single pixle event    & 155298 & 54.73$\%$     \\
        2-pixel split event    & 74808  & 26.36$\%$     \\
        3-pixel split event    & 41881  & 14.76$\%$      \\
        4-pixel split event & 11787 & 4.15$\%$      \\ \hline
        \end{tabular}
\end{table}

\subsection{Alignment with ``Burkert test'' method}\label{X-ray imagination}
% To confirm the X-ray imaging capability of the detector, a 3D printed plastic mask with the letter ``IHEP eXTP'' is installed on the cylindrical baffle to conduct an X-ray imaging experiment. 
% The size of the letter pattern is about 20 mm $\times$ 20 mm. 100 frames are acquired with a characteristic line of Fe, and then the images are subtracted by dark field image with the same exposure condition to subtract the blemish luminescence. 
% 100 frames of X-ray image are superimposed, as shown in Fig.~\ref{CMOS_image}, proving the X-ray imaging ability of the focal plane detector. 

% \begin{figure}[htb]
% 	\begin{center}
% 	    \includegraphics[width=0.6\textwidth]{camera_mask.pdf}
% 		\includegraphics[width=0.6\textwidth]{CMOS_image.pdf}
% 		\caption{Top panel: The layout of the imaging experiment. 
% 		The mask with the letter ``IHEP eXTP'' is placed in front of the focal plane detector. 
% 		Bottom panel: The photo of the 3D printed plastic mask under X-ray exposure.
% 		The blemish luminescence is subtracted through dark field images.}
% 		\label{CMOS_image}
% 	\end{center}
% \end{figure}

\begin{figure}[htb]
	\begin{center}
		\includegraphics[width=0.6\textwidth]{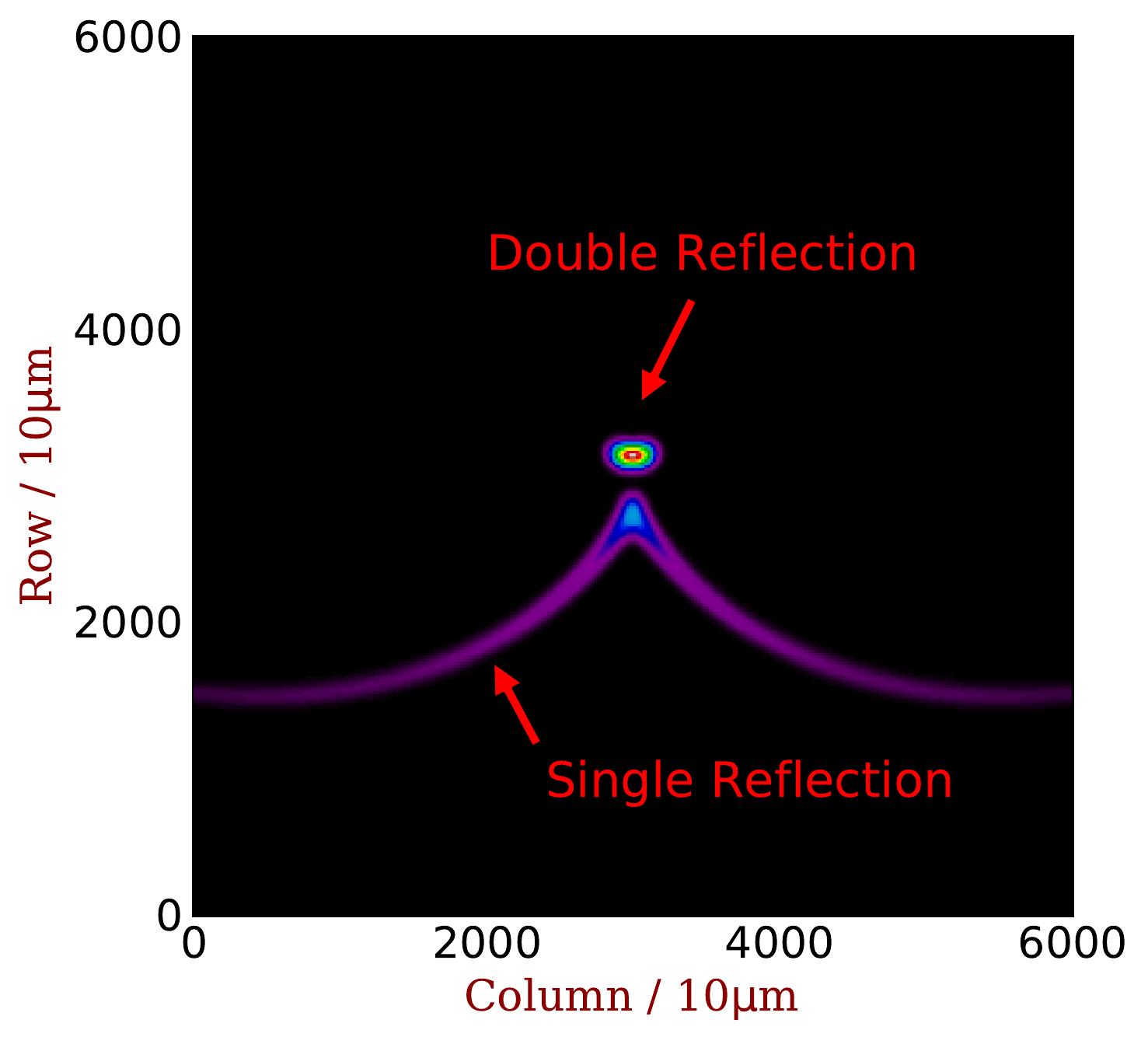}
		\caption{Image of optical simulation with single Wolte-\RNum{1} type X-ray mirror when applying ``Burkert test'' method. 
		The off-axis pitch angles is set to 0.3$^{\circ}$.}
		\label{off_axis_off_focus_simulation}
	\end{center}
\end{figure}

The X-ray mirror modules for SFA and PFA are under development, hence, the process of aligning the optical axis of single Wolter-\RNum{1} type X-ray mirror with the X-ray beam optical axis with ``Burkert test'' method is simulated based on optical model. 
At the focal point, the off-axis pitch angles of the mirror is set to 0.3$^{\circ}$.
The single reflection and double reflection X-ray imaging with sCMOS sensor GSENSE6060BSI are shown in Fig.~\ref{off_axis_off_focus_simulation}. 
Therefore, an imaging detector with an area of 60 mm $\times$ 60 mm meets the requirements of the ``Burkert test'' method.

\begin{figure}[htb]
	\begin{center}
		\includegraphics[width=0.6\textwidth]{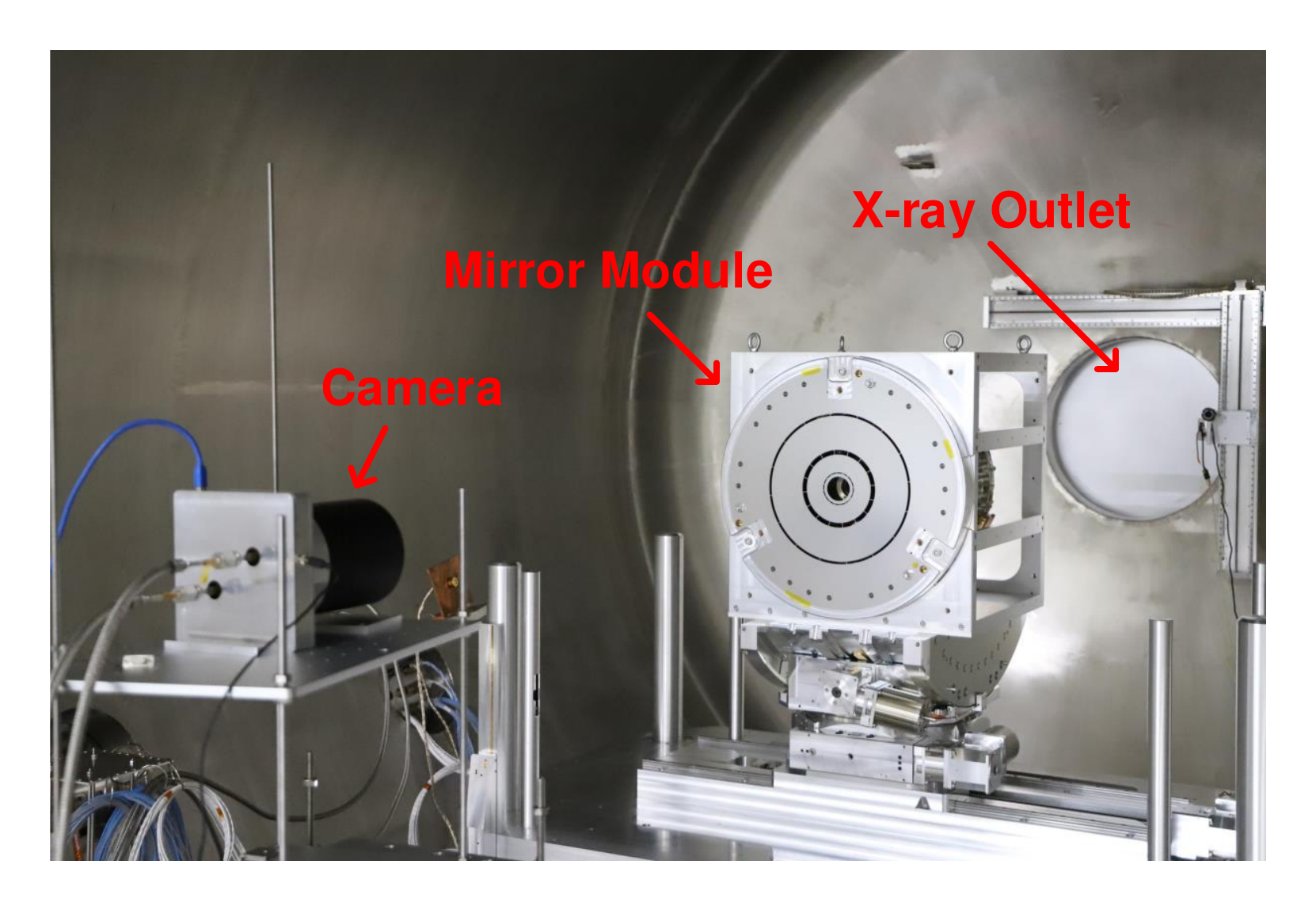}
		\caption{
		Setup of the mirror module and the focal plane detector in a vacuum chamber.
		The camera and the mirror module are placed approximately on the same optical axis with an X-ray outlet.
		Only 4 selected mirror shells of the module are used by installing a mask behind the mirror module.
		}
		\label{sCMOS_FXT_mirror_test_setup}
	\end{center}
\end{figure}

\begin{figure}[htb]
	\begin{center}
	    \includegraphics[width=0.6\textwidth]{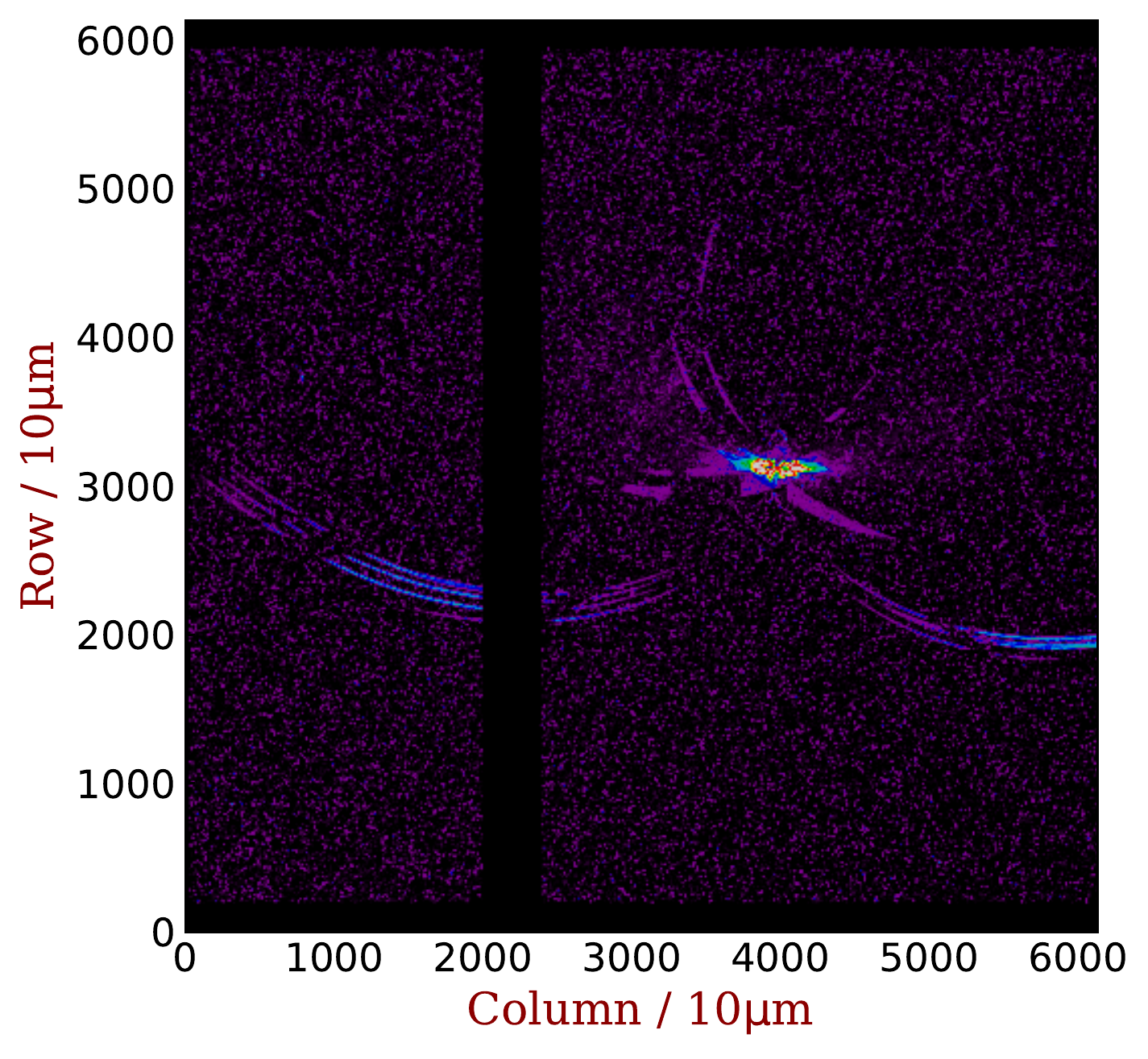}
		\includegraphics[width=0.6\textwidth]{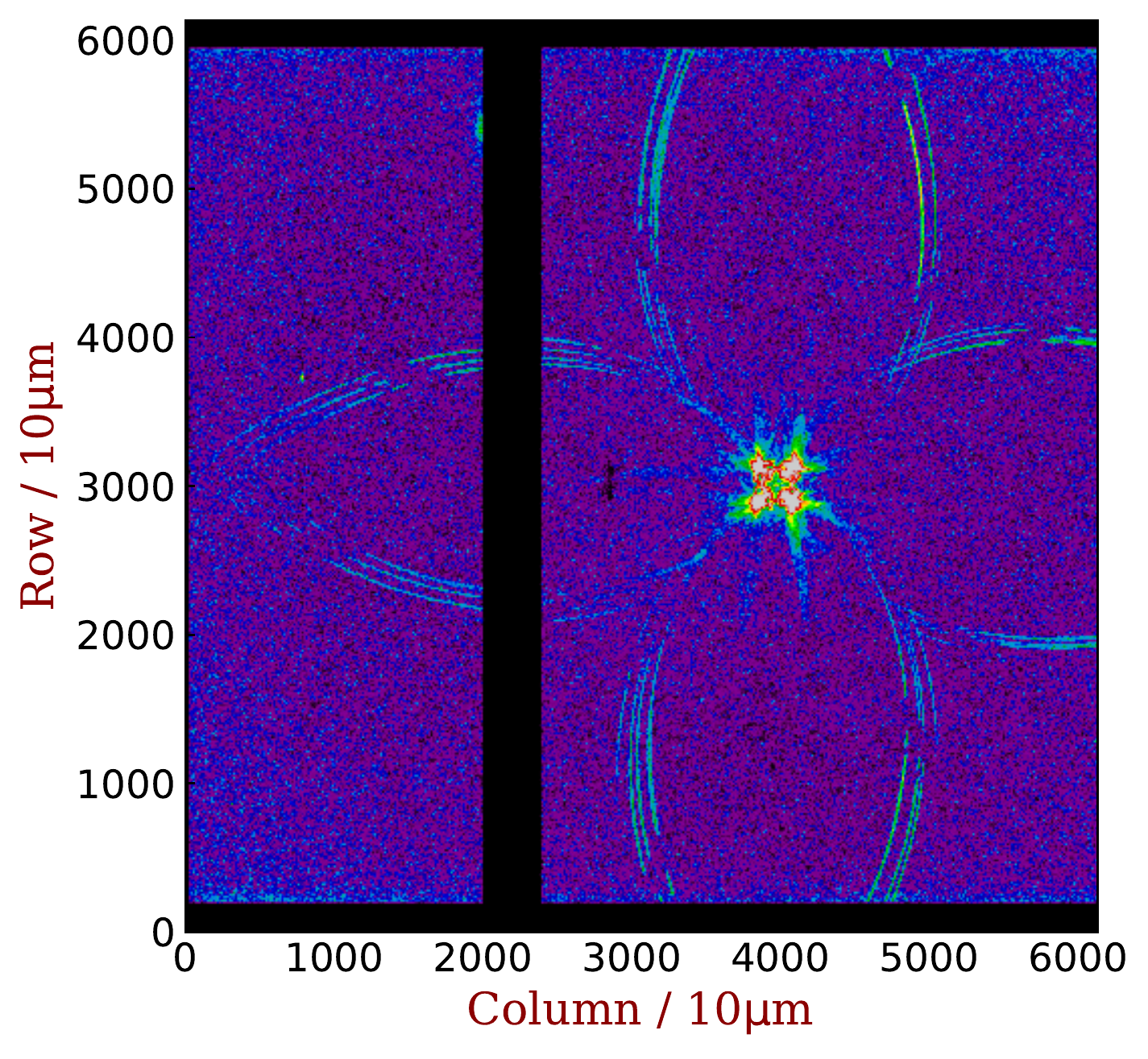}
		\caption{
		The upper panel is the image of single reflection light with the off-axis pitch angle of 0.7$^{\circ}$, which is consist with the simulation shown in Fig.~\ref{off_axis_off_focus_simulation}. 
		The bottom panel is the combined image of four single reflections. 
		The off-axis pitch and yaw angles are set to $\pm$0.7$^{\circ}$.
		The data from 2000th to 2500th column and 0 to 200th, 5944th to 6144th rows are deleted to reduce the influence of the blemish luminescence and diffusion luminescence, respectively.
		}
		\label{FXT_mirror_single_reflection}
	\end{center}
\end{figure}

The mirror module designed for \textit{EP/FXT} (Einstein Probe/Follow-up X-ray Telescope)\cite{EP-FXT} is used to verify the feasibility of X-ray mirror alignment through the ``Burkert test'' method. 
Fig.~\ref{sCMOS_FXT_mirror_test_setup} shows the test setup of the mirror and the focal plane detector in vacuum chamber. 
Part of the data is deleted to reduce the influence of the blemish luminescence and diffusion luminescence\cite{JamesCCD} on the single reflection light, as shown in Fig.~\ref{FXT_mirror_single_reflection}. Four off-axis single reflection lights are combined and plot in one canvas, which is consistent with the simulation results in Fig.~\ref{off_axis_off_focus_simulation}. 
The reflected light at each off-axis yaw and pitch angle is roughly symmetrical, which provides a rough scan range for a finer alignment method\cite{bradshaw2019developments}. 

\section{Conclusion and discussion}\label{conclusion}
A large area focal plane detector is developed based on the scientific back-illuminated sCMOS sensor GSENSE6060BSI produced by \textit{Gpixel Inc.}, which meets the requirements for eXTP mirror module testing.
According to the requirements of vacuum environment and thermal control, a corresponding optimization design is taken into account, and modular design is adopted to make the detector easy to upgrade in the future. 

The detector is tested with a multi-target fluorescence X-ray source to study the readout noise and dark current characteristics, energy detection lower limit, energy linearity, and characteristics of the split-pixel event. 
The readout noise and dark current are about 3.2 $\rm e^-$, 18 $\rm e^-/pixel/s$, respectively.
We can conclude from the experiment that the low limit of energy detection is about 186 eV, and the characteristic line of Fe-$\rm L{\alpha}$ (0.7 keV) is obtained through event reconstruction. 
The integral nonlinearity (INL) is about 0.35$\%$ in the energy range of 1.2 keV to 8.9 keV. 
Finally, the optical simulation and alignment experiment of Wolter-\RNum{1} type X-ray mirror with the ``Burkert test'' method are carried out.

At present, more and more micro and small satellites in the field of X/$\gamma$-ray astronomy research have begun or are being prepared to choose CMOS sensor as a focal plane detector, including the under-development EP \cite{yuan2018einstein}, HiZ-GUNDAM \cite{2020HiZ-GUNDAM}, etc., hence, the performance of GSENSE6060BSI demonstrated in this research is of great significance in sCMOS sensor selection and its performance evaluation for the following satellites.

\section{Acknowledgments}
This study is supported by the Strategic Priority Program on Space Science, the Chinese Academy of Sciences, Grant No. XDA15020501 and No. XDA1531010301. 
We also thank for the valuable discussion with Zhixing Ling from the National Astronomical Observatories, Chinese Academy of Sciences.

% \paragraph{Note added.} This is also a good position for notes added
% after the paper has been written.


\bibliography{reference}

\end{document}